# Kerr Frequency Combs: A Million ways to fit light pulses into tiny rings

Aurélien COILLET[1*], Shuangyou ZHANG[2], Pascal DEL'HAYE[2]

[1] *Laboratoire Interdisciplinaire Carnot de Bourgogne, Dijon, France*
[2] *Max Planck Institute for the Science of Light, Erlangen, Germany*

* *aurelien.coillet@u-bourgogne.fr*

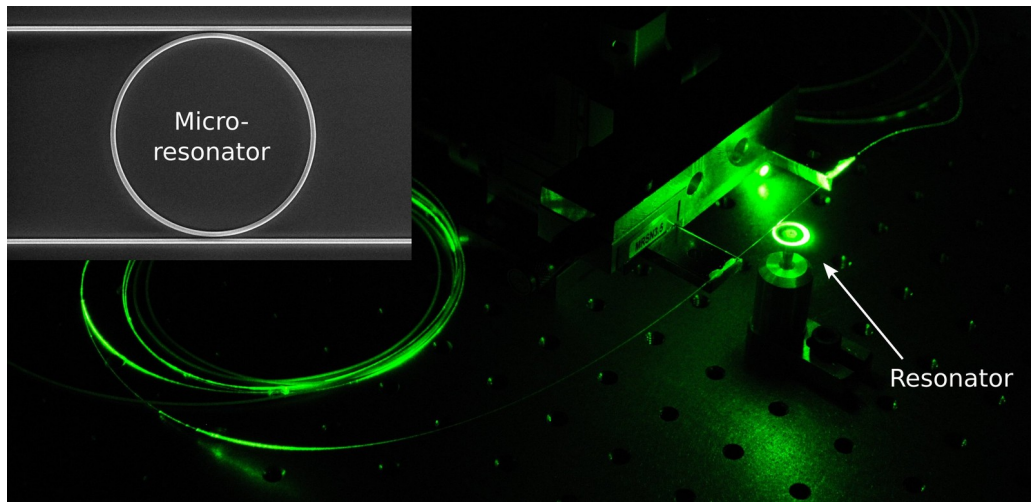

*Frequency combs can be generated in millimeter-sized optical resonators thanks to their ability to store extremely high light intensities and the nonlinearity of their materials. New frequencies are generated through a cascaded parametric amplification process which can result in various optical waveforms, from ultrastable pulse patterns to optical chaos. These Kerr frequency combs have been studied extensively, with a wealth of fascinating nonlinear dynamics reported, and myriads of applications being developed, ranging from precision spectroscopy and Lidars to telecom channel generators.*

1. Introduction

    Kerr frequency combs are a novel way to generate frequency combs based on the nonlinear interaction between a single wavelength light source and a dielectric material [1]. Since such interactions are quite weak, a resonator which forces light to recirculate in the material is used, accumulating sufficient optical power until new frequencies are generated through a nonlinear process called four-wave mixing (FWM, see insert). Depending on the parameters of both the resonator and pump light used, a variety of optical patterns can be generated and used for different applications ranging from time-and-frequency metrology to building blocks for integrated photonic circuits. The temporal dynamics of the combs generation in microresonators can be described by the Lugiato-Lefever equation (a nonlinear Schrödinger equation), which predicts exceptionally well the dynamics of these systems.

2. Recipe for a Kerr frequency comb

The first and most important ingredient for Kerr comb generation is undoubtedly the resonator itself. It needs to have a relatively large Kerr nonlinearity, but most importantly, extremely low losses, so that light can remain trapped for a long time. Typically, its quality factor Q – which is proportional to the lifetime of a photon in the cavity – has to be larger than $10^6$ to cross the threshold for parametric oscillations at reasonably low optical power. The next step consists in coupling as much monochromatic light as possible into the resonator using the evanescent field of a microfiber, cleaved fiber, waveguide or a prism, and scan the input frequency to excite one resonance. When the pump light within the resonator exceeds the threshold for comb generation, new frequencies are generated. The generated comb (corresponding to solitons in the time domain) is coupled out of the resonator for further analysis or for use in applications.

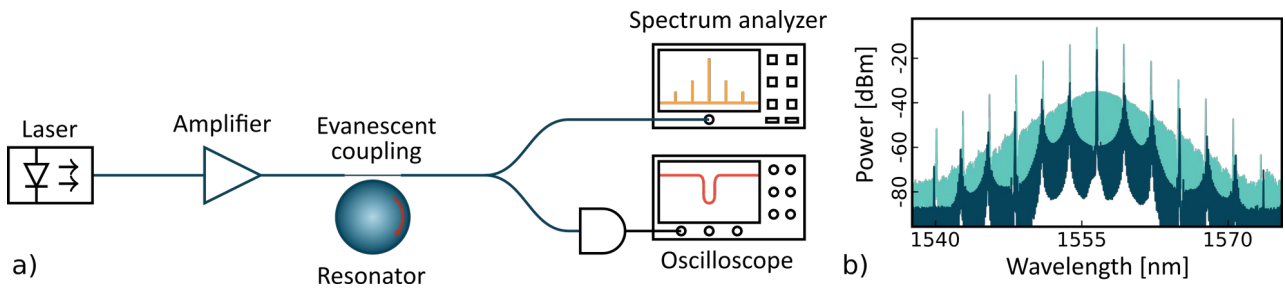

**Figure 1.** a) Simplest experimental scheme for Kerr comb generation with an amplified, single-wavelength laser source coupled to a resonator using a microfiber in this case. The output signal can directly be monitored through the same coupling microfiber. b) Examples of different Kerr combs (light and dark blue) generated in the same resonance, at the same power, but for a slightly different detuning between the cavity resonance and the pump frequency.

Depending on its exact frequency compared to the resonance, the pump laser light will be more or less enhanced, and all of the newly generated frequencies will interact differently with the resonances of the resonator, leading to a wide variety of circulating soliton pulse patterns.

3. The Lugiato-Lefever equation

The initial observation of various types of Kerr combs lead to the search for an appropriate theoretical model that could describe the experimental observations. One can understand Kerr comb generation either by looking at it as a collection of spectral modes that interact through four-wave mixing [2], or rather like the propagation of light in a closed-loop, with a continuous pump wave constantly pouring energy into the system [3]. In both of these approaches, one finds that the optical field in the cavity follows the following normalized equation:

$$\frac{\partial \psi}{\partial \tau} = -(1+i\alpha)\psi + i|\psi|^2\psi - i\frac{\beta}{2}\frac{\partial^2 \psi}{\partial \theta^2} + F$$

The evolution of the field is hence governed by the interactions between losses, the frequency detuning between the pump and the resonance $\alpha$, the normalized nonlinearity, the second order dispersion of the resonator $\beta$ and the pump field amplitude $F$. With these notations, the field $\psi$ depends both on the long timescale time $\tau$ and the angle $\theta$ which marks the position within the resonator.

This equation is a modified version of a nonlinear Schrödinger equation, with both losses and driving included, and is known as the Lugiato-Lefever equation. This equation was introduced in 1987 by Luigi Lugiato and René Lefever [4] in order to model the spontaneous formation of patterns in 2D resonant Kerr media, and a large corpus of theoretical analyses is already available. In particular, the analysis of the stability of the flat solution, that is when no other frequencies or combs are generated, predicts in which region of the parameter space of interesting structures can be formed [5]. For our purpose, a resonator has its dispersion fixed during the manufacturing process, and the only 2 parameters that remains accessible to the experimentalist are the detuning of the laser with respect to the resonance frequency $\alpha$ and the pump power $F^2$.

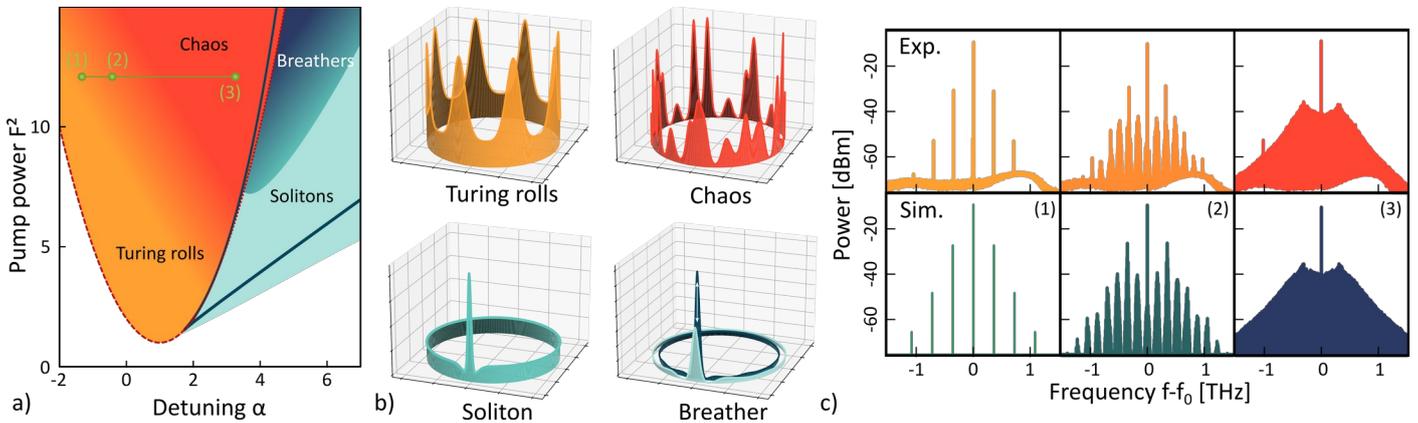

**Figure 2.** a) Bifurcation diagram of the 1D Lugiato-Lefever equation in the anomalous dispersion regime ($\beta<0$) with the regime of Kerr comb one can expect in each region of the parameter space. The detuning $\alpha$ is the difference between laser and resonance frequency normalized by the resonance bandwidth. b) Numerical simulations of the intracavity intensity for the most notable regimes of Kerr comb. Turing rolls and soliton are stable structures, while breather and chaotic regimes vary with time. c) Comparison between experiments and the LLE model for the spectra of the generated Kerr combs, while varying the detuning along the green line in a).

In the anomalous dispersion regime, a wealth of comb states can be achieved: Turing rolls appear spontaneously above a given threshold, and correspond to sine-like oscillation of the intensity. A soliton is obtained when the laser frequency is slightly higher than the resonance frequency (positive detuning), but requires a seed pulse or significant fluctuations to be generated, as this part of the parameter space is multi-stable; multiple excitations can therefore lead to multiple solitons. Increasing the pump power can lead to instability, with breathers and chaotic pulsed regimes taking place.

Despite its mathematical simplicity, the Lugiato-Lefever has been able to model the experimental results obtained with various resonators with an impressive precision, as can be seen in Figure 2 c). The evolution from a stable comb to chaos is reproduced with high accuracy over several orders of magnitude of intracavity power. Such an accuracy stems from the very simple processes at play, and particularly the parametric gain: the Kerr effect is indeed quasi-instantaneous, with a very large spectral bandwidth, such that its modelling is very easy.

4. Beyond the Lugiato-Lefever equation

Rich cavity soliton states have been predicted by Lugiato-Lefever equation and experimentally observed in different dispersion regimes and with different pumping schemes. As shown in Fig. 3, with a single frequency pumping and anomalous dispersion, bright soliton pulses (intensity peaks on a dark background) can be generated through a well-defined balance between the Kerr nonlinearity, group velocity dispersion, cavity loss and gain [6]. Crystallized soliton structures with crystallographic defects have also been observed in this regime. These correspond to fixed patterns of solitons that are stable within the resonator and repeat after each round-trip. In contrast, the normal dispersion regime leads to dark pulses (intensity dips embedded in a high-intensity background) that arise through the interlocking of switching waves connecting the homogeneous steady states of the bi-stable cavity solutions (middle panel in the right of Fig. 3) [7]. Distinct soliton structures in the zero-dispersion regime have also been reported with asymmetrical behaviour both in the frequency domain and time domain. The zero-dispersion regime is of particular interest because it allows to investigate how higher-order dispersion affects the soliton formation dynamics. The soliton (bright or dark) formation in this regime is associated with the interlocking of abrupt changes of power in a resonator, so-called switching waves. Most importantly, zero or small dispersion is very desirable to obtain spectrally broadband frequency combs. In addition, the coexistence and mutual locking of dark and bright pulses in the regime of normal, zero, and anomalous dispersion has been predicted by the Lugiato-Lefever equation when considering higher-order dispersion effects. Recent research revealed bound states of dark-bright solitons in a resonator through seeding two modes with opposite dispersion via two colors of light (lower panel in the right of Fig. 3) [8]. These mutually trapped dark-bright pulses lead to a light state with constant output power in the time domain but spectrally resembling frequency combs.

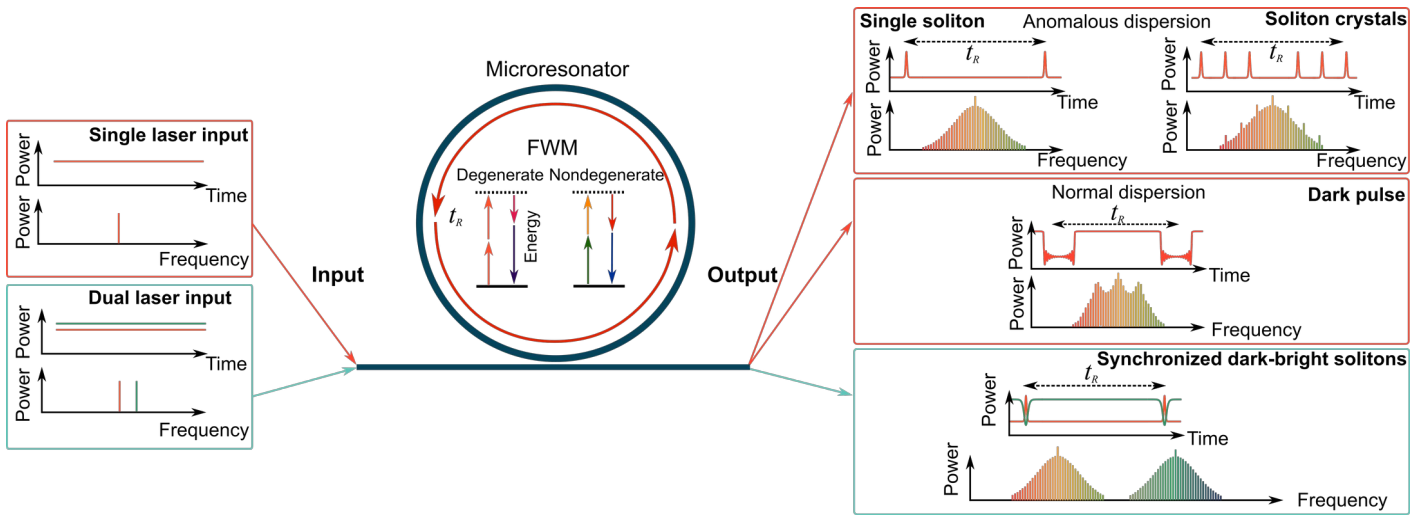

**Figure 3.** Different light pulses emitted from the resonators. With single frequency driving of a resonator (marked with red boxes), bright solitons and soliton crystals can be generated in anomalous dispersion regime, while dark pulses can be observed by driving at normal dispersion. Seeding with two colors of lights, dark-bright soliton bound states can be emitted from a resonator with a mutually trapped dark-bright soliton pair.

In addition to the generation by a CW laser source, cavity solitons can be generated with even higher efficiency by synchronously pulsed-driving. The corresponding soliton dynamics are also well modelled by the Lugiato-Lefever equation. An additional way for the generation of Kerr solitons is the use of active media as part of an external cavity, which has been studied with increasing attention in the last years. This concept has the advantage of not requiring a narrowband tuneable laser source for the soliton generation.

5. Conclusion

Frequency combs generated through four-wave mixing in high quality factor resonators constitute a very promising alternative to mode-locked lasers for many applications where compacity is required. In the past 15 years, research on this topic went from early demonstration of frequency generation to theoretical understanding and prediction, and to applications, including optical frequency synthesis, ultrastable microwave generation, calibration for astronomy spectroscopy, and ranging. Most of these applications make use of a single pulse revolving inside the cavity – a unique soliton – yet a myriad of other exotic and physic-rich regimes can be obtained. The Lugiato-Lefever equation and its refinements have been used to accurately predict the generation of these original states, such as soliton crystals and locked switching waves. The full extent of the possibility offered by Kerr frequency combs and their variants – pulse-driven, dual-pump, with gain medium, … – remains to be explored, both for fundamental study in nonlinear dynamics and for demanding applications.

___________________________________________________________________________________

Insert 1.

Four-wave mixing is one of the effects of the third-order Kerr optical nonlinearity: assuming two optical waves with different frequencies $\nu_1$ and $\nu_2$ are co-propagating inside a Kerr material, a modulation occurs which lead to the creation of two new frequencies $\nu_3$ and $\nu_4$ such that energy is preserved, that is $\nu_1 + \nu_2 = \nu_3 + \nu_4$. One can also use on single pump wavelength, as is the case for Kerr frequency comb generation, and such process is then called degenerate four-wave mixing. Once new frequencies have been generated though, a cascading effect can occur and all the spectral lines can interact through four-wave mixing. Since the Kerr effect is rather low in most materials, a high power is required for this conversion to be efficient. In the case of Kerr frequency comb generation, such high power is reached thanks to the cavity enhancement of the optical field at resonance. Like most Kerr effects, four-wave mixing is phase-sensitive, hence leading to inherently phase-locked frequency lines.

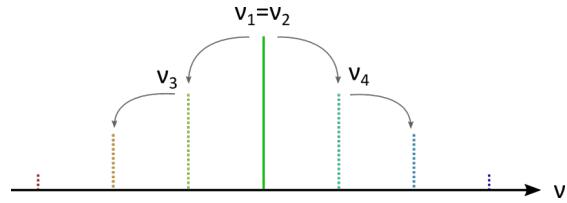

___